\documentstyle[aps,multicol,epsf,psfig]{revtex}
\newcommand{\ket}[1]{| #1 \rangle}

\begin{document}

\title{Minimal Absorption Measurements}
\author{ Serge Massar$^{1}$, Graeme Mitchison$^{2}$ and Stefano Pironio$^{1}$}
\address{$^{1}$Service de Physique Th\'eorique,
Universit\'e Libre de Bruxelles, CP 225, Bvd. du Triomphe, B1050
Bruxelles, Belgium.\\
$^{2}$MRC Laboratory of Molecular Biology, Hills Road,
Cambridge CB2 2QH, UK.\\
(February 22, 2001) }

\maketitle

\begin{abstract}

In this paper we consider the problem of trying to make an image of an
object while minimizing the number of photons absorbed by the
object. We call protocols which achieve this goal ``minimal
absorption measurements''. Such imaging techniques are particularly
relevant in situations where the object can be damaged by the
radiation used to make the image. Our main results are bounds that
relate the minimum number of absorbed photons to the sensitivity of
the measurement. In the case where the object consists of a single
pixel, we show that these bounds can be approached either by
simply counting the number of photons absorbed by the object or by a
simple interferometric setup, depending on the details of the
problem. In the case where the object consists of many pixels, we give
an example where our bound can be approached when all the pixels are
addressed collectively, whereas addressing each pixel individually
implies an increase in the number of absorbed photons by a factor
logarithmic in the number of pixels.  Finally, we consider some
special situations where our bound does not apply, and where
interferometric methods can make large gains.

\end{abstract}
{ PACS numbers: 03.65.Bz}\\
\pacs{03.65.Bz}

\vspace{-1.1cm}


\section{Introduction}\label{one}

We examine the possible advantages of using the quantum properties of
light for making images of photosensitive objects, i.e. objects that
can be damaged by the photons that are used to make the image.  The
recent advances in quantum information processing suggest that using
the quantum nature of light may provide advantages over classical
imaging techniques.  In particular Elitzur and Vaidman\cite{EV} have
proposed ``interaction-free measurements'' (generalized in
\cite{KwiatEtal95}) that can determine whether a completely absorbing
object is present or absent with infinitesimal probability that the
object absorbs a photon.  Absorption-free measurements (as we prefer
to call them, rather than ``interaction-free'', since our
terminology reduces the number of inverted commas) have been further
generalized in \cite{1} to the case where the object is
semi-transparent. And in \cite{KSS} it was shown that an
interferometric setup is useful when one wants to determine
simultaneously the probability that a semitransparent photosensitive
object absorbs a photon and the phase it induces on a photon that
traverses it but is not absorbed (i.e. one wants to measure both the
real and imaginary part of the transparency $\alpha$ introduced
below).

Thus quantum mechanics may provide a powerful new way of making
images of photosensitive objects, and this has motivated various
experimental implementations of absorption-free measurements
\cite{KwiatEtal95,exp}. However, some doubt has been cast on the wider
applicability of these methods by the finding that it is impossible to
distinguish unambiguously between two semitransparent objects (neither
of which is totally transparent) without a certain non-zero
probability that the objects absorb a photon \cite{1}.

In this note we explore a different problem from that considered in
\cite{1} which is probably more important in practice.  We consider
the case where one wants to determine the transparency of an object
with high resolution while {\em minimizing} the number of photons
absorbed by the object. We shall show in this case that, except for
certain specialised tasks, using the quantum mechanical properties of
light does not offer significant advantages over more traditional
schemes such as counting absorbed photons or simple interferometric
procedures that use one photon at a time.

The general scenario that we have in mind is an object which is
characterized by a position-dependent transparency
$\alpha(x)$. Here $x$ is a coordinate on the surface of the object,
and $\alpha$ is the complex amplitude for a photon not to be absorbed
by the object. The aim is determine $\alpha(x)$ to high resolution
while minimizing the number of photons absorbed by the object. A
``minimal absorption measurement'' is a protocol whose outcome (a set
of measurement results) achieves this aim.

This scenario is difficult to analyze in full generality, and hence we
simplify it. As a first step, we consider an extreme simplification in
which (a) the object consists of only one pixel (i.e. $x$ takes 
only one value), and (b) there are only two possible
objects. That is, the transparency of the single pixel can take only
two values, $\alpha_1$ or $\alpha_2$. The task is then to determine
which object is present (ie. to determine whether the transparency is
$\alpha_1$ or $\alpha_2$) while absorbing as few photons as
possible. We give a bound on the minimum number of photons that must
be absorbed if objects are to be distinguished with high
probability. This bound is valid for an arbitrary quantum protocol.
In the conclusion we mention how this result can be extended to the
case where the transparencies can take a continuous range of values.

To state our result in the single pixel case, let
$\alpha=(\alpha_1+\alpha_2)/2$ be the average of $\alpha_1$ and
$\alpha_2$, and let $\beta_1$, $\beta_2$ be the amplitudes that
objects 1 and 2 absorb a photon.  Let $\beta = \sqrt{1-|\alpha|^2}$,
so $\beta$ is approximately the average of $|\beta_1|$,
$|\beta_2|$. We are interested in the case where the difference of
transparencies is small, so we can write $\alpha_1 = \alpha -
\epsilon$ and $\alpha_2 = \alpha + \epsilon$, where $\epsilon$ is a
small (complex) number satisfying $|\epsilon| \ll {\beta^2 \over 2
|\alpha|}$. Let $P_E$ denote the probability of making a mistake in
identifying the object.  Denote by $\bar N_{i}^{abs}$ the mean number of
absorbed photons in a minimal absorption measurement if object $i$ is to be
correctly identified with probability greater than $1-P_E$.  Then our
main result is the following constraint on $\bar N_{1}^{abs}$ and
$\bar N_{2}^{abs}$:
\begin{equation} 
{\bar N_{1}^{abs} + \bar N_{2}^{abs} \over 2} \geq {\beta^4 (1-2
\sqrt{P_E(1-P_E)}) \over 2 |\epsilon|^2} + O(1) \ .
\label{E1}
\end{equation}

The most important aspect of (\ref{E1}) is that the mean number of
absorbed photons increases as $1/|\epsilon|^2$. In most cases this
bound is comparable to the resolution that can be obtained with very
simple schemes which do not exploit the full range of possibilities
offered by quantum mechanics.  Indeed simply counting the number of
photons absorbed by the object gives a resolution comparable to the
bound of (\ref{E1}), except in the special case where $\alpha_1$ and
$\alpha_2$ differ only in phase.

When $\alpha_1$ and $\alpha_2$ differ only in phase one must use an
interferometric protocol. A simple interferometric protocol, in which
the photons can either pass through the object or not do so,
yields a $1/\epsilon^2$ dependence of the mean number of absorbed
photons, as in   (\ref{E1}).  However we have only been able to
attain the same fourth power dependence on $\beta$ as in
(\ref{E1}) by modifying the interferometric protocol in such a way
that the photon can either pass a large number $k$ of times through
the object or not pass through it at all.

As a further step, we consider the situation where one aims to
distinguish not just the transparencies of one pixel but different
{\it patterns} of transparency $\alpha(x)$ on a set of pixels. We
obtain in this case a bound on the minimum number of photons that must
be absorbed if objects are to be distinguished with high probability
which is similar to the bound obtained in the single pixel case. 

We illustrate the multi-pixel situation by an example of a special
discrimination task, where a significant decrease in the number of
absorbed photons can be obtained by using interferometric protocols
that address all the pixels simultaneously, as compared to addressing
each pixel individually.  This example is based on recent work of Wim
van Dam\cite{vanDam}. All transparencies are assumed to be real,
i.e. the pixels differ only by the probability that they absorb a
photon. We show that a collective measurement of all the pixels
decreases the number of absorbed photons, relative to measurements on
pixels one by one, by a factor of $\log(\mbox{\emph{Number of
pixels}})$. Similar decreases in the number of absorbed photons when
imaging a multi-pixel object have   been obtained independently by
Adrian Kent and David Wallace\cite{KW}.

Finally, we consider some examples which illustrate the limitations of
our results. One example is provided by absorption-free measurements in the
case $\alpha_1=1$ where small differences can be resolved with negligible
photon absorption. A second is based on Grover's algorithm \cite{G}, which
allows essentially absorption-free discrimination in some special situations.
We explain why these examples escape the limitations of our bound (\ref{E1}).

The paper is organized as follows. The next three sections are
concerned with the case of a single pixel and only two
objects. Specifically sections \ref{two} and \ref{three} present
protocols that yield photon-absorption numbers of the same order as
our bound (\ref{E1}), and section \ref{four} contains the proof of
this bound. Sections \ref{five} and \ref{six} are concerned with the
general case of a multi-pixel image and many possible objects. Section
\ref{five} obtains an analogous bound to inequality (\ref{E1}) in this
more general case, and section \ref{six} describes our example in
which collective addressing yields a decrease in the number of
absorbed photons. Section \ref{seven} discusses the limitations of our
approach, with some examples, and section \ref{eight} contains our
conclusions and discusses possible extensions of our results.

\section{Counting the number of absorbed photons}\label{two}

In this section and the next two, we consider the case where we have a
single pixel, and the transparency of this pixel can take only one of
two values, $\alpha_1$ or $\alpha_2$.  Let us suppose that one
sends, one by one, $N$ photons through the unknown object and counts
how many pass through it without being absorbed.  For object \( i \)
(\( i=1,2 \)), this number is distributed according to a binomial of
mean $\mu_i = |\alpha_i|^2 N$ and standard deviation $\sigma \simeq
|\alpha \beta| \sqrt{N}$. The decision strategy that minimizes the
probability of error is to compute for which object the observed
number of transmitted photons (denoted by $N^T$) is most likely, and
to guess that that is the correct object.  Thus there will be some
value $N^{T*}$ (with $\mu_1 \leq N^{T*} \leq \mu_2$) such that if
$N^T\leq N^{T*}$ one guesses that $i=1$ and if $N^T>N^{T*}$ one
guesses that $i=2$ (we have taken $|\alpha_1| <|\alpha_2|$).

For small error probability $P_E$, the number of photons sent through
the object $N$ must be large and the two binomials will tend to
Gaussians. $N^{T*}$ is then approximately equal to $\mu_1 + \mu_2
\over 2$ and the probability of error is approximately $P_E\simeq {1
\over 2}(1-\mbox{erf}({\mu_2-\mu_1 \over \sqrt 8 \sigma}))$. This gives
$(\mu_2-\mu_1)/2\sigma=\sqrt2\gamma(P_E)$, where $\gamma
(P_E)=\mbox{erf}^{-1}(1-2P_E)$ (so, for instance, $\gamma=0.91$ when
$P_E=0.1$, or $\gamma=1.65$ when $P_E=0.01$). Hence the mean number of
absorbed photons for this protocol is
\begin{equation}
\bar N^{abs} = \beta^2 N \simeq {2  \beta^4 
|\alpha|^2 \gamma(P_E)^2 \over (\alpha \bar\epsilon+\bar \alpha \epsilon)^2 }.
\label{nc1}
\end{equation}

In the above procedure we have supposed that the photons are sent one
by one and that we have a perfect single photon detector at our
disposal. We could also have sent a state containing $N$ identical
photons and measured the number of transmitted photons. This supposes
that one can prepare photon number states and measure photon number.
In practice it is of course much easier to send as input a coherent
state of amplitude $A$ and measure the intensity transmitted through
the object. In this case one finds that $\bar N^{abs} \geq \beta^2
|\alpha|^2 \gamma'(P_E)^2/(\alpha \bar\epsilon+\bar \alpha
\epsilon)^2$ where $\gamma'(P_E)$ is the analogue of $\gamma$ for a
Poisson distribution. It is interesting to note that these two
procedures differ in the value of the pre-factor ($\beta^4$ or
$\beta^2$), the lesser efficiency of using coherent states presumably
being due to the uncertain photon number for such states. We note
that since all the other protocols discussed in this article use
linear optical elements (beam splitters, phase shifters) they can also 
be implemented using coherent light. In these cases one also expects
an increase in the mean number of absorbed photons for the reason just 
mentioned.

Returning to inequality (\ref{nc1}), when $\alpha$ and $\epsilon$ are
real this reduces to
\begin{equation}
\bar N^{abs} \simeq {\beta^4 \gamma(P_E)^2 \over 2\epsilon^2},
\label{nc2}
\end{equation}
so in this real case $\bar N^{abs}$ is proportional to
$1/|\epsilon|^2$, as in (\ref{E1}). For complex $\alpha$ and
$\epsilon$, the denominator in (\ref{nc1}) may be much smaller than
$|\epsilon|^2$; in particular, if $|\alpha_1|=|\alpha_2|$, so
$\alpha_1$ and $\alpha_2$ differ only in phase, or equivalently if
$\alpha \bar\epsilon+\bar \alpha \epsilon=0$, then $\bar
N^{abs}=\infty$, and photon-counting fails. In this situation we need
an interferometric procedure that is sensitive to the phase of
$\alpha$ and $\epsilon$.

\section{Interferometric procedures}\label{three}

We focus on the case where $|\alpha_1|=|\alpha_2|$, so simple counting of
absorbed photon fails. In this case, one can write  $\alpha_{1,2}=\alpha e^{
\pm i\eta}$, where $\alpha$ is real (any global phase can be removed by an
appropriate phase shifter placed on the path of the photon sent
through the object). Note that $|\beta|^2 = 1 - \alpha^2 = 1 -
|\alpha_{1,2}|^2$ since $
\alpha_{1,2}$ differ only by a phase.
For $\eta$ small, this can be rewritten as $\alpha_{1,2}\simeq
\alpha \pm i \alpha \eta = \alpha \pm i \epsilon$, where $\epsilon = \alpha
\eta $.

A simple interferometric procedure to detect object 1 or 2 consists of
a Mach-Zender interferometer with the object located in one of the
arms. The transmission and reflection amplitudes at the first
beam-splitter are \( 1/\sqrt{2} \), while at the second beam-splitter
the transmission amplitude is
\( 1/\sqrt{2} \) and the reflection amplitude \( i/\sqrt{2} \). A photon
originally in the state \( \ket {0} \) is sent to \( (\ket {0}+\ket
{1})/\sqrt{2} \) by the first beam-splitter.  This state then becomes \( (\ket
{0}+\alpha _{i}\ket {1})/\sqrt{1+|\alpha _{i}|^{2}} \) through interaction with
the object. Finally, after the second beam-splitter, the probability of
detecting the photon in arm \( 0 \) with object $i$ is \( \chi_i =|1+i\alpha
_{i}|^{2}/2(1+|\alpha _{i}|^{2}) \).

The idea is to send \( N \) photons one by one through the
interferometer, count the number of times the detection of a photon in
arm \( 0 \) occurs, and to distinguish objects \( 1 \) and \( 2 \) by
these counts. The mean number of measurement outcome is \( \mu
_{i}=N\chi_i \), so that \( \mu _{2}-\mu _{1} \simeq 2N\epsilon
/(1+\alpha^{2}) \). The standard deviation of the distribution of
these counts is \( \sigma \simeq \sqrt{N\chi (1-\chi)} \), where \(
\chi = |1+i\alpha|^{2}/2(1+\alpha^{2})=1/2 \). Applying the
same criterion for distinguishability as in section \ref{two}, one
finds \( \bar{N}^{abs}=\gamma ^{2}\beta ^{2}(1+\alpha
^{2})/2\epsilon^{2} \), so \( \bar{N}^{abs} \) is again proportional
to \( 1/|\epsilon |^{2} \).

Note, however, that there is a factor of $\beta^2$ missing, compared
to the   bound, $\bar N^{abs} \geq O(\beta^4/|\epsilon|^2)$ of
(\ref{E1}), and this means that the above simple interferometric
method diverges from the multi-photon bound as $\alpha$ tends to 1 and
$\beta$ to zero.  To recover the correct order in $\beta$, we simply
let the photon pass $k$ times through the object before the second
beam-splitter instead of just once. The probability of detecting it in
arm 0 is then \( \chi_i =|1+i(\alpha _{i})^k|^{2}/2(1+|\alpha
_{i}|^{2k}) \). Counting the number of times detection occurs gives
$\mu_2-\mu_1 \simeq 2Nk\alpha^{k-1} \epsilon/(1+\alpha^{2k})$ and
$\chi \simeq 1/2$. We are interested in the
case where $\alpha$ is close to 1. Writing $\alpha=1-\delta $, with
$\delta$ small, and taking $k=1/\delta$, we find $\mu_2-\mu_1 \simeq
2N \epsilon/\delta(e+e^{-1})$. The total absorption probability is
${1\over 2}(1-\alpha^{2k}) \rightarrow {1 \over 2} (1-e^{-2})$.  Thus
$\bar N^{abs} \simeq O(\delta^2/\epsilon^2)$, and since
$\beta^2=1-(1-\delta)^2 \simeq 2\delta$ this gives $\bar N^{abs}
\simeq O(\beta^4/|\epsilon|^2)$, which is the same order as the
  bound, (\ref{E1}), in both $\epsilon$ and $\beta$.

Thus we can use simple counting of absorbed photons to achieve the
quantum bound, except when $\alpha_1$ and $\alpha_2$ differ only in
phase, in which case an interferometric procedure yields a mean
number of absorbed photons of the same order as the bound
(\ref{E1}). We now turn to the proof of inequality (\ref{E1}).

\section{Bounds for Minimal Absorption Measurements}\label{four}

Absorption-free measurement schemes use a single photon and a choice
of paths for that photon: through the arm of the interferometer where
the object is located, or through the other `ancillary' arm. However,
these protocols are not the most general ones.  Indeed, if a single
photon is absorbed the protocols stop. But since we are not
considering absorption-free measurements but rather minimal absorption
schemes we should allow for the possibility that some of the photons
may be absorbed while the remaining photons are still available for
further operations. We can then imagine protocols where we send a
superposition of states with different photon numbers through the
object, and possibly perform collective measurements on these
photons. Or we can use ancillas which do not interact with the object
to create entangled state such as $| \mbox{\emph{2 photons not passing
through the object}} \rangle | \mbox{\emph{0 photons passing through
the object}} \rangle + | \mbox{\emph{0 photons not passing through the
object}} \rangle | \mbox{\emph{2 photons passing through the object}}
\rangle$, and all sorts of combinations of this kind.

In order to give a bound on the efficiency of these quantum protocols,
we must give a completely general formulation of such protocols. To
this end, we divide the total Hilbert space into the product of three
subspaces \( H_{A}\otimes H_{P}\otimes H_{O} \).  The first subspace,
\( H_{A} \), defines the state of the ancilla. It can be arbitrary;
for instance it can be a subspace of the Fock space of the
electromagnetic field. The second subspace, \( H_{P} \), describes the
photons which are sent through the object. A basis of \( H_{P} \) is
therefore the photon number states $\ket{n}_P$.  The last subspace, \(
H_{O} \), corresponds to the state of the object.  If \( n_{1},\ldots
n_{j},\ldots \) photons have been absorbed by the object at stages
$1, \ldots,j , \dots$ of the protocol, the state of the object
becomes \( \ket {n_{1},\ldots ,n_{j},\ldots}_{O} \) and represents the
state induced in the object by the absorptions. That part of the
object Hilbert space that is of interest to us can therefore be
written as the tensor product of Fock spaces: \( H_{O}^{1}\otimes
H_{O}^{2}\otimes \cdots \), where number states in \( H_{O}^{j} \)
count the number of photons that have been absorbed at stage $j$.
The object can also have internal degrees of freedom, which we do not
write explicitly.

We now adopt the general formulation of a protocol used in
\cite{1}. It was assumed there that a protocol starts from a specified
initial state, after which there is a succession of steps, called
`interaction steps', where the photons are sent through the object,
alternating with steps where some (arbitrary) unitary transformation
occurs. Finally some measurement is made which seeks to determine
which object was present. In our general minimal absorption
protocol, we assume there is an initial state of the ancilla and
photons, \( \ket{\Psi_0}_{AP} \), with the object in state \( \ket
{0_{1},0_{2},\ldots }_{_{O}} \) since no interaction has yet taken
place. The unitary transformation following the $j$-th interaction
step is of the form $U_{AP}^j\otimes I_O$.

In \cite{1} the interaction step is assumed to consist of two parts: a
unitary transformation which describes the initial interaction between
photon and object, and a measurement which describes the subsequent
collapse of the photon/object system into `interacted' or
`non-interacted' outcomes (the object is assumed to be
macroscopic). Here we assume that each interaction step consists only
of a unitary transformation, and we do not carry out the collapse
step. If we were to trace over the object space $H_O$ at the end of
the protocol, before the final measurement, this would be equivalent
to the complete interaction step in \cite{1}. Instead, however, we
assume for mathematical simplicity that the final measurement is
completely arbitrary and can also act on the state of the object. This
can only increase the information gain and hence the efficiency of the
protocol. For this reason the bound given by inequality (\ref{E1}) is
probably not optimal.

We assume the interaction step takes the form of a unitary
transformation $I_A\otimes U^{int}_{PO}$, where $I_A$ is the identity
on the ancilla Hilbert space. The action of $U^{int}_{PO}$ is given by
\begin{equation}
\label{int}
a^{\dagger }_{_{P}}\rightarrow \alpha a^{\dagger }_{_{P}}+\beta
b^{\dagger }_{j_{O}},
\end{equation}
where \( a^{\dagger }_{_{P}} \) and \( b^{\dagger }_{j_{O}} \) are the
creation operators in \( H_{P} \) and \( H^{j}_{O} \),
respectively. The evolution that this induces on the state \( \ket
{a}_{_{A}}\ket {1}_{_{P}}\ket {0_{j}}_{_{O}}=a^{\dagger }_{_{P}} \ket
{a}_{_{A}}\ket {0}_{_{P}}\ket {0_{j}}_{_{O}} \), which represents a
single photon sent through the object (and an ancilla), is\[ \ket
{a}_{_{A}}\ket {1}_{_{P}}\ket {0_{j}}_{_{O}}\rightarrow \alpha \ket
{a}_{_{A}}\ket {1}_{_{P}}\ket {0_{j}}_{_{O}}+\beta \ket {a}_{_{A}}\ket
{0}_{_{P}} \ket {1_{j}}_{_{O}}.\] If \( l \) photons are sent through
the object, the unitary evolution (\ref{int}) gives
\begin{equation}
\label{lphot}
\begin{array}{lcl}
\ket {a}_{_{A}}\ket {l}_{_{P}}\ket {0_{j}}_{_{O}}=\frac{(a^{\dagger
    }_{_{P}})^{l}}{\sqrt{l!}}\ket {a}_{_{A}}\ket {0}_{_{P}}\ket
{0_{j}}_{_{O}} & \rightarrow  & \frac{(\alpha a^{\dagger
    }_{_{P}}+\beta b^{\dagger }_{j_{O}})^{l}}{\sqrt{l!}}\ket
{a}_{_{A}}\ket {0}_{_{P}}\ket {0_{j}}_{_{O}}\\ 
 &  & =\sum _{m=0}^{l}(\begin{array}{c}
l\\
m
\end{array})^{1/2}\alpha ^{m}\beta ^{l-m}\ket {a}_{_{A}}
\ket {m}_{_{P}}\ket {(l-m)_{j}}_{_{O}}\\
 &  & =\ket {a}_{_{A}}\ket {\widetilde{l}_{j}}_{_{PO}}
\ .
\end{array}
\end{equation}
In the last line we have introduced the notation
\begin{equation}
\label{postabs}
\ket {\widetilde{l}_{j}}_{_{PO}}=\sum _{m=0}^{l}(\begin{array}{c}
l\\
m
\end{array})^{1/2}\alpha ^{m}\beta ^{l-m}\ket {m}_{_{P}}\ket {(l-m)_{j}}_{_{O}}
\end{equation}
that will serve us later.

We use the following notation for the state function, given
transparency $i$, during successive stages of the protocol:
\[
\ldots \ket{\Psi^i_j} \to \mbox{Interaction Step}_j \to \ket{\Phi^i_j} \to \mbox{Unitary}_j \to  \ket{\Psi^i_{j+1}} \to \mbox{Interaction Step}_{j+1}  \to \ket{\Phi^i_{j+1}} \ldots \mbox{etc.}
\] 
With this notation, we can explicitly write the state immediately
before the $j$-th interaction step as
\begin{eqnarray}
\ket {\Psi ^{i}_{j}}_{_{APO}}=\sum _{k,l,n_{1},\ldots
  n_{j-1}}C^{j,i}_{kln_{1},\ldots n_{j-1}}\ket {k}_{_{A}}\ket
{l}_{_{P}}\ket {n_{1},\ldots n_{j-1},0_{j},0_{j+1},\ldots }_{_{O}}
\nonumber
\end{eqnarray}
where \( \{\ket {k}_{_{A}}\; ,\: k=0\ldots S\} \) is a basis in \(
H_{A} \) and \( \{\ket {l}_{_{P}}\} \) are the Fock states in \(
H_{P} \).

Immediately after the interaction step, \( \ket {\Psi
^{i}_{j}}_{_{APO}} \) becomes
\begin{eqnarray}
\ket {\Phi ^{i}_{j}}_{_{APO}}&=&
I_A\otimes U^{int}_{PO} \ket {\Psi ^{i}_{j}}_{_{APO}} \nonumber\\
&=&
\sum _{k,l,n_{1},\ldots
  n_{j-1}}C^{j,i}_{kln_{1},\ldots n_{j-1}}\ket {k}_{_{A}}\ket
{\widetilde{l}^{i}_{j}}_{_{PO}}\ket {n_{1},\ldots
  n_{j-1},0_{j+1},\ldots}
\nonumber
\end{eqnarray}
where we have used the notation of (\ref{postabs}).

After this interaction step, the unitary transformation $U_{AP}\otimes
I_O$ transforms \( \ket {\Phi ^{i}_{j}}_{_{APO}} \)to \( \ket {\Psi
^{i}_{j+1}}_{_{APO}} \).  Following \cite{1}, we define
\[
f_{j}=|\langle \Psi _{j}^{1}|\Psi ^{2}_{j}\rangle |.
\]
The overlap $f_{j}$ plays an important part in our argument. It
measures how much the two states corresponding to evolution with the
two transparencies $i=1,2$ differ, and thus how easily one can
distinguish them. The smaller this quantity, the more efficient the
protocol.

Unitarity implies that
\[
f_{j+1}=|\langle \Phi _{j}^{1}|\Phi ^{2}_{j}\rangle |,\]
which, using the normalization conditions
\[
\begin{array}{l}
_{_{A}}\langle k|k^{'}\rangle _{_{A}}=\delta _{kk^{'}}\\
_{_{O}}\langle n_{1}\ldots n_{j-1}|n^{'}_{1}\ldots n^{'}_{j-1}\rangle
_{_{O}}=\delta _{n_{1}n^{'}_{1}}\ldots \delta_{n_{j-1}n^{'}_{j-1}}\, , 
\end{array}\] 
can be written as
\begin{equation}
\label{prodsc}
f_{j+1}=|\sum _{k,l,l^{'},n_{1},\ldots
  n_{j-1}}(\overline{C}_{l}^{j,1}C^{j,2}_{l^{'}})_{kn_{1}\ldots
  n_{j-1}}\, _{_{PO}}\langle
\widetilde{l}^{1}_{j}|\widetilde{l}^{'2}_{j}\rangle _{_{PO}}| \ . 
\end{equation}
Using (\ref{postabs}), we can compute \( _{_{PO}}\langle
\tilde{l}_{j}^{1}|\tilde{l}'^{2}_{j}\rangle _{_{PO}} \): 

\[
_{_{PO}}\langle \tilde{l}_{j}^{1}|\tilde{l}'^{2}_{j}
\rangle _{_{PO}}=\delta _{ll'}\: \sum _{m=0}^{l}(\begin{array}{c}
l\\
m
\end{array})(\bar \alpha _1\alpha _2)^{m}(\bar{\beta
  }_{1}\beta _{2})^{l-m}=\delta _{ll'}\: (\bar{\alpha }_{1}\alpha
_{2}+\bar{\beta }_{1}\beta _{2})^{l}\ .
\] We can then rewrite
(\ref{prodsc}) as
\begin{equation}
f_{j+1}=|\sum _{k,l,n_{1},\ldots n_{j}}
(\bar{C}^{j,1}C^{j,2})_{kln_{1}\ldots n_{j}}
(\bar{\alpha }_{1}\alpha _{2}+\bar{\beta }_{1}\beta _{2})^{l}|\ .
\label{absum}
\end{equation}

We now obtain an approximation for the term $\bar\alpha
_1\alpha_2+\bar\beta_1\beta_2$. The assumption we are making that the
final measurement can act on the state of the object is physically
incorrect. We should be tracing over the Hilbert space of the object
before the final measurement, in which case the phases of the
absorption amplitudes $\beta_i$ would be irrelevant. We are therefore
free to choose the phases of the $\beta_i$ as we wish, so as to obtain
the best bound. Suppose therefore that
\[
\bar\beta_1\beta_2=e^{i\phi}\sqrt{1-\bar\alpha_1\alpha_1}
\sqrt{1-\bar\alpha_2\alpha_2}, 
\]
where $\phi$ is real but otherwise can be chosen freely. By adjusting
$\phi$ appropriately, with the assumption $|\epsilon| \ll \beta^2$, we
can make $\bar\alpha_1\alpha_2+\bar\beta_1\beta_2$ real. Indeed, we
can rewrite $\bar\alpha_1\alpha_2+\bar\beta_1\beta_2$ as
$e^{i(\rho_1-\rho_2)}\cos\theta_1\cos
\theta_2+e^{i\phi}\sin\theta_1\sin\theta_2$, where $\cos \theta_i
=|\alpha_i|$ and where $\rho_i$ is the phase of $\alpha_i$. The
imaginary terms vanish if
$\sin\phi=\cos\theta_1\cos\theta_2\sin(\rho_2-\rho_1)/(\sin\theta_1\sin\theta_2)$. This
is possible only if the right hand side is $\leq 1$, which is ensured
if $|\epsilon| < {\beta^2 \over 2 |\alpha|}$.

With this choice of $\phi$, we find
\begin{equation}
\bar \alpha_1\alpha_2+\bar\beta_1\beta_2 = 1-{2|\epsilon|^2 \over \beta^2} +
O(\epsilon^4)
\label{devlop}
\end{equation}
Introducing the notation
\[
(\bar{\alpha }_{1}\alpha _{2}+\bar{\beta }_{1}\beta
_{2})^l = 1- \delta_l,
\]
eq (\ref{devlop}) implies that
\[
\delta_l \leq l\left ({2|\epsilon|^2 \over \beta^2} +
O(\epsilon^4)\right )
\]
since $(1-x)^l\ge 1-lx$, for $0 \le x \le 1$ (as can easily
be verified by induction on $l$).

Turning back to eq (\ref{absum}) we are now able to compute $f_{j+1}$:
\begin{equation}
\label{grineq}
\begin{array}{rl}
f_{j+1} & =|\sum _{k,l,n_{1},\ldots
 n_{j}}(\bar{C}^{j,1}C^{j,2})_{kln_{1}\ldots n_{j}}(\bar{\alpha
 }_{1}\alpha _{2}+\bar{\beta }_{1}\beta _{2})^{l}|\\ 
& =|\sum _{k,l,n_{1},\ldots
  n_{j}}(\bar{C}^{j,1}C^{j,2})_{kln_{1}\ldots n_{j}}(1-\delta_{l})|\\
& \geq |\sum _{k,l,n_{1},\ldots
  n_{j}}(\bar{C}^{j,1}C^{j,2})_{kln_{1}\ldots n_{j}}|-|\sum
_{k,l,n_{1},\ldots n_{j}}(\bar{C}^{j,1}C^{j,2})_{kln_{1}\ldots
  n_{j}}\delta _{l}|\\ 
& \geq f_{j}-\sum _{k,l,n_{1},\ldots
  n_{j}}|(\bar{C}^{j,1}C^{j,2})_{kln_{1}\ldots n_{j}}\delta _{l}|\\ 
& = f_{j}- ({2 |\epsilon| ^{2} \over \beta^2} + O(\epsilon^4))
\sum _{k,l,n_{1},
\ldots n_{j}}|(\bar{C}^{j,1}C^{j,2})_{kln_{1}\ldots n_{j}}|l\\
 & \geq f_{j}-({2 |\epsilon| ^{2} \over \beta^2} + O(\epsilon^4))
\sum _{k,l,n_{1},\ldots
   n_{j}}\frac{|C^{j,1}_{kln_{1}\ldots n_{j}}|^{2}+|C_{kln_{1}\ldots
     n_{j}}^{j,2}|^{2}}{2}l\\ 
 & =f_{j}-({2 |\epsilon| ^{2} \over \beta^2} + O(\epsilon^4))
(n_{j}^{1}+n_{j}^{2})/{2}
\end{array}
\end{equation}

In the last equality, \( n_{j}^{i}= \sum _{k,l,n_{1},\ldots
n_{j}}|C^{j,i}|^{2}_{kln_{1}\ldots n_{j}}l \) is the average number of
photons sent through the object at interaction step \( j \).  Starting
from \( f_{1}=1, \) and iterating the formula above, we conclude that
\begin{equation}
\label{Ksum}
f_{K}\geq 1-({2 |\epsilon| ^{2} \over \beta^2} + O(\epsilon^4)) \sum
_{j=1}^{K-1}\frac{n_{j}^{1}+n_{j}^{2}}{2}=1- ({2 |\epsilon| ^{2} \over
\beta^2} + O(\epsilon^4)) \frac{N^{1}+N^{2}}{2},
\end{equation}
\( N^{i} \) being the total number of photons that are sent through
object \( i \) in the protocol. The average number of photons
absorbed by object $i$ is $\bar N_i^{abs}=|\beta_i|^2 N^i$, so that
\[
{\bar N_{1}^{abs} + \bar N_{2}^{abs} \over
2}={|\beta_1|^2N^1+|\beta_2|^2N^2 \over 2}\simeq \beta^2 {N^1+N^2
\over 2}
\]
where the last approximation is valid if $|\epsilon| \ll {\beta^2 \over 2
|\alpha|}$. From (\ref{Ksum}), we thus find that
\begin{equation}
{\bar N_{1}^{abs} + \bar N_{2}^{abs} \over 2} \geq {\beta^4
  \over 2 |\epsilon|^2} (1-f_K) + O(\epsilon^0).
\end{equation}

By choosing measurement axes symmetrical with respect to the final
state vectors $\Psi^1_K$ and $\Psi^2_K$, one obtains the minimal
probability of mis-identification of the object, $P_E ={1 \over
2}(1-\sqrt{1-f_K^2})$, which implies $f_K=2\sqrt{P_E(1-P_E)}$
\cite{Hels}. Putting everything together gives our   bound
(\ref{E1}).

\section{Discriminating faint images}\label{five}

We now consider the more general situation where we have $M$ pixels,
and $L$ images, i.e. assignments of transparencies to these pixels.
The transparencies at any one pixel for all the different images are
assumed to be close. We obtain a bound on the mean number of absorbed
photons that is similar to inequality (\ref{E1}), with $\epsilon$
replaced by the maximum difference between the transparencies of
individual pixels. Because this bound does not decrease with the
number of pixels, it is not obvious that collective addressing of the
pixels offers a significant advantage over addressing each pixel
individually. Nevertheless we shall give in the next section an
example where such collective addressing significantly decreases the
mean number of absorbed photons.

The state at stage $j$ when there are $M$ pixels and $L$ images can be
written as
\[
\ket{\psi^p_j}_{APO}= \sum C^{j,p}_{k,l_1,\ldots,l_M; n_{11},n_{12},
  \ldots , n_{1M},n_{21},\ldots , n_{j-1 M}} \ket{k}_A \ket{l_1}
\ldots \ket{l_M} \ket{n_{11},\ldots}, 
\]
where $p$, $1 \le p \le L$ specifies the image, $\{\ket
{k}_{_{A}}\; ,\: k=0\ldots S\}$ is a base in $H_{A}$, $l_i$ is the
number of photons sent through pixel $i$ and $n_{ab}$ is the number of
photons absorbed at stage $a$ by pixel $b$. We can write the
coefficient concisely as $C^{j,p}_{k,\bf l,n}$.

If $f^{p,q}_j$ denotes the absolute value of the overlap between
states with image $p$ and $q$ at stage $j$, the analogue of
eq. (\ref{absum}) is
\[
f^{p,q}_{j+1}=\left| \langle \psi^p_{j+1} | \psi^q_{j+1} \rangle \right| =\left| \sum_{k,\bf l,n}  \left(\bar C^{j,p}C^{j,q}\right)_{k,\bf l,n} \prod_{i=1}^M(\bar\alpha^p_i\alpha^q_i+\bar\beta^p_i\beta^q_i)^{l_i} \right|,
\]
where $\alpha^p_i$ is the amplitude for absorbing a photon at pixel
$i$ given image $p$. Assuming that
$\epsilon^{p,q}_i=(\alpha^p_i-\alpha^q_i)/2$ is small, for any $i$ and
$p$, $q$, we can write this approximately as 

\[
f^{p,q}_{j+1} \simeq \left| \sum \left(\bar C^{j,p}C^{j,q}\right)_{k,\bf l,n}
(1-\sum_i \delta^i_{l_i}) \right|, 
\]
where $\delta^i_l \le l(2|\epsilon^{p,q}_i|^2/\beta_i^2 +
O((\epsilon^{p,q}_i)^4))$, and $\beta_i$ is the average of
$|\beta_i^p|$ over all images $p$. Leaving out the $O(\epsilon^4)$
terms, the steps leading to eq. (\ref{grineq}) and (\ref{Ksum}) then
yield
\[
\sum_{k,\bf l,n} \sum_{i=1}^M
{|C^{j,p}_{k,\bf l,n}|^2+ |C^{j,q}_{k,\bf l,n}|^2 \over
2}l_i{2|\epsilon^{p,q}_i|^2
\over \beta_i^2} \ge 1-f^{p,q}_K.
\]
or 
\begin{equation}
\sum^M_{i=1} (N_i^p+N_i^q){|\epsilon^{p,q}_i|^2 \over \beta_i^2} \ge
1-f^{p,q}_K, 
\label{manybetas}
\end{equation}
where $N_i^p$ is the number of photons passing through pixel $i$ under
assignment $p$. Now consider the special situation (which will arise
in the next section) where the $\alpha^p_i$ are close in value for all
$i$ and $p$, i.e. over all pixels and all images. Let $\beta$ be the
average of all the $|\beta^p_i|$s, and let $|\epsilon^{p,q}|=\max_i
\{|\epsilon^{p,q}_i|\}$a. Then we have the corollary to (\ref{manybetas})
\begin{equation}
\label{eq:n}
\bar N^{abs}_p + \bar N^{abs}_q \ge {\beta^4  (1-f^{p,q}_K) \over
  |\epsilon^{p,q}|^2} 
\end{equation}
where $\bar N^{abs}_p$ is the mean number of photons absorbed if the
image is $p$. In the case where we wish to discriminate between two
images, $p$ and $q$, we obtain the analogue of inequality (\ref{E1}):
\begin{equation}
\label{eq:multiE1}
{\bar N^{abs}_p + \bar N^{abs}_q \over 2} \ge {\beta^4 ( 1 - 2
 \sqrt{P_E(1-P_E)}) \over 2|\epsilon^{p,q}|^2},
\end{equation}
where $P_E$ is the probability of mistaking image $p$ for the image
$q$. If the task is to distinguish one image from all $L-1$
others, then averaging over all $p \ne q$ gives
\begin{equation}
\label{eq:hat}
{\hat  N^{abs}} \ge {\beta^4 ( 1 - 2
 \sqrt{P_E(1-P_E)}) \over 2|\epsilon|^2},
\end{equation}
where $|\epsilon|=\max_{p,q}\{|\epsilon^{p,q}|\}$, $P_E$ now
denotes the maximum probability of confusing any two images, and $\hat
N^{abs}$ is the mean absorption $\bar N^{abs}_p$ averaged over all
images $p$.

\section{Logarithmic gain in the number of absorbed photons}
\label{six}
 
The bound (\ref{eq:multiE1}) is simply the single-pixel bound
(\ref{E1}) applied to the pixel $i$ for which the difference of
transparencies $|\epsilon_i^{p,q}| = |\alpha_i^p - \alpha_i^q| / 2$ is
largest.  This bound does not depend on the number of pixels in the
image. Nevertheless we shall show that it is possible to decrease the
number of absorbed photons by addressing all the pixels collectively, as
compared to addressing them one by one.

  Wim van Dam \cite{vanDam} considered the following problem:
Given a Hadamard matrix $H$, an oracle evaluates the function
$f(i)=H_{pi}$, where $H_{pi}$ is the $i$-th element of the $p$-th row
of $H$, for some specified $p$. One wishes to know the smallest number
of calls to the oracle needed to determine $p$. We consider an
analogous problem here. Suppose $\alpha^p_i=\alpha+\epsilon H_{pi}$,
for small $\epsilon$. The task is to determine $p$ with a minimum
number of photons absorbed. We can think of the $\epsilon H_{pi}$ term
as defining a faint image against the background of $\alpha$. This is
an `epsilon' version of van Dam's problem, with a passage of a photon
through the pixels playing the role of a call to an oracle. We assume
that $\alpha$ and $\epsilon$ are real.
 
Suppose the Hadamard matrix is un-normalized, so the entries are $ \pm
1$, and that the top row of $H$ is all $+1$s. Given two rows, $p$ and
$q$, $p \ne q$, $\epsilon^{p,q}_i=0$ at the pixels $i$ where the rows
$p$ and $q$ agree, and $\epsilon^{p,q}_i=\epsilon$ otherwise. Thus
inequality (\ref{eq:hat}) implies
\[
\hat N^{abs} \ge {\beta^4 (1-f) \over 2 \epsilon^2},
\]
and $\hat N^{abs}$ is $O(1/\epsilon^2)$.
 
Following \cite{vanDam},
one can achieve this bound with the following simple quantum
algorithm. Assume $p \ne 1$. Prepare the one photon state
$\ket{\psi}={1 \over \sqrt M} \sum \ket{i}$. After the photon passes
through the pixels once, there is a probability $\alpha^2+\epsilon^2$
of the photon not being absorbed, in which case $\ket{\psi}=\sum
(\alpha+\epsilon
H_{pi})\ket{i}/\sqrt{M(\alpha^2+\epsilon^2)}$. Applying the unitary
operator $H/\sqrt{M}$ to this gives
$(\alpha\ket{1}+\epsilon\ket{p})/\sqrt{\alpha^2+\epsilon^2}$, and
$\ket{p}$ can be detected with $O(1/\epsilon^2)$ independent repeats
of the process. Thus we get the same order as our bound.

Let us now suppose that we address each pixel individually.
Specifically we shall suppose that we send photons one by one through
the pixels and measure whether they are absorbed or not.  The
following informal argument gives a bound on the value for $\bar
N^{abs}$ for such an individual-pixel protocol. Let $A=\{p_A,p_N\}$ be
the distribution with probability $p_A$ of absorption and $p_N$ of
non-absorption at pixel $i$. Let $R=\{p_1, \dots ,p_M\}$ be the
distribution of probabilities of the Hadamard rows. The mutual
information $I(A:R)=H(A)+H(R)-H(A,R)$ tells us how much on average we
learn (in bits) on being told whether a photon is absorbed at pixel
$i$ or not.

Given prior probabilities $\pi_p$ for row $p$, $H(R)=-\sum_p \pi_p
\log \pi_p$. The term $H(A,R)$ is given by
\[
H(A,R)=-\sum_p \{P(A,p) \log P(A,p)+P(N,p) \log P(N,p)\},
\]
where $P(X,p)$ is the joint probability of row $p$ being chosen and
the photon being absorbed ($X=A$) or not ($X=N$). Since
$P(A|p)=(\alpha+H_{pi}\epsilon)^2$, we have
$P(A,p)=P(A|p)P(p)=(\alpha+H_{pi}\epsilon)^2 \pi_p$. Similarly,
$P(N,p)=\{1-(\alpha+H_{pi}\epsilon)^2\}\pi_p$. Finally, $H(A)=-p_A
\log p_A-p_N \log p_N$, where $p_X=\sum_p P(X,p)$, for $X=A, N$. Up to
second order in $\epsilon$ one finds
\[
I(A:R) \simeq {2\epsilon^2 \over \beta^2 \log 2} \{\sum_p
\pi_pH_{pi}^2-(\sum_{p} \pi_p H_{pi})^2\} \le
{2\epsilon^2 \over \beta^2 \log 2}.
\]

As $\log_2 M $ bits are needed to distinguish one row from $M$ others,
and each photon yields at most $4\epsilon^2/(\beta^2 \log 2)$ bits,
we need at least $N$ photons, where $4N \epsilon^2/(\beta^2 \log 2)
\ge \log_2 M$, or $N \ge \beta^2 \log M/(4\epsilon^2)$. Thus
$N^{abs} \ge \beta^4\log M/(4\epsilon^2)$. The $\log M$ factor here
is the counterpart of the $\log n$ classical bound given in
\cite{vanDam}, Lemma 5.

A value of $N^{abs}$ of this order can be attained by an algorithm
in which the same number $N/M$ of photons is sent through each
pixel. Denoting by $n_i$ the number of photons transmitted at pixel
$i$, the algorithm seeks the value of $q$ which maximizes the sum
$S_q=\sum_i H_{qi}n_i$ if $q \neq$ top row, $S_q=\sum_i
H_{qi}n_i-N\alpha^2$ if $q=$ top row.  It is easy to check that the
distributions for $S_q$ have standard deviation $\alpha \beta\sqrt N$
and the difference of the means for $p$ and $q \ne p$ is $2N\alpha
\epsilon$. To distinguish one distribution from amongst $M$ we
therefore need $\alpha \beta \sqrt{N \log M}=2N \alpha \epsilon$,
implying $\bar N_{abs}=O(\log M/\epsilon^2)$. Thus the collective and
individual-pixel bounds can both be attained, and there is a
collective gain of a factor of $\log M$.

\section{Beating the bounds: limitations of our results.}\label{seven}

We now give some examples where it is possible to make large gains
over simple counting of absorbed photons. These examples may seem to
violate our main inequality (\ref{E1}), but we explain why this is not
so.

Consider first absorption-free measurement \cite{EV}. Given a
completely absorbing and a completely transparent object, this offers
the possibility of discrimination without any photons being absorbed
(with probability approaching one \cite{KwiatEtal95}). Of course, our
inequality (\ref{E1}) does not apply here, since the transparencies
being compared are not close in value. However, in the case where
$\alpha_1=1$, and $\alpha_2$ takes some value not equal to $\alpha_1$,
essentially absorption-free discrimination is always possible
\cite{1}. Suppose we choose $\alpha_2=1-2\epsilon$ for small
$\epsilon$. Then $\beta \simeq \sqrt{2\epsilon}$, and, carrying
through the calculations leading to (\ref{E1}), one finds $N^{abs} \ge
(1-2\sqrt{P_E(1-P_E)})(1-\epsilon)$.  Since the absorption-free
measurement gives an arbitrarily small $N^{abs}$ with $P_E=0$, we seem
to have a violation of our inequality. However, there is in fact no
contradiction, since the condition $|\epsilon| \ll {\beta^2 \over 2
|\alpha|}$ is not satisfied, and this is used in the proof of (\ref{E1}).

It is interesting to note that the inequality in \cite{1} for the
probability of absorption-free discrimination yields the same
order dependence as (\ref{E1}) in the range where they may
legitimately be compared. In fact, if one asks how many photons
would be absorbed by repeating an absorption-free protocol until it
succeeds, one can express this in terms of the probability $P(abs|i)$
of a photon being absorbed during the protocol. One sums over repeated
runs with absorptions until no absorption occurs, when the object will
be reliably identified. Thus
\begin{equation}
\bar N_i^{abs}=\sum_{k=0}^{\infty} k (1-P(abs|i))P(abs|i)^k= {P(abs|i) \over (1-P(abs|i))}
\end{equation}
We know from \cite{1} that $P(abs|1)P(abs|2) \geq \eta^2$ where
$\eta={|\beta_1\beta_2| \over |(1-\bar \alpha_1 \alpha_2)|}$. With the same
assumption as in our proof of (\ref{E1}), i.e. $|\epsilon| \ll {\beta^2 \over
2 |\alpha|}$, we find ${|\beta_1\beta_2| \over |(1-\bar \alpha_1 \alpha_2)|}
\simeq 1- {2|\epsilon|^2 \over \beta^4}$. There is thus necessarily a value of
$i$ so that $P(abs|i)\geq\eta$. This implies that for that value of $i$ we have
\begin{equation} \bar N_i^{abs} \geq {\eta \over 1-\eta} \simeq {\beta^4 \over
2|\epsilon^2|}, \end{equation} with the same dependence on $\beta$ and
$\epsilon$ as our bound (\ref{E1}).

We turn now to another example, where collective measurement of many
pixels apparently enables our bound to be beaten. This example is
inspired by Grover's search algorithm\cite{G}. Recall that the oracle
in Grover's algorithm carries out the transformation $\ket{x}\ket{y}
\to \ket{x}\ket{y \oplus f(x)}$ where $x=1,\ldots, M$ is the position,
$y=0,1$ and $f(x)=0$ except if $x=x_0$ in which case $f(x_0)=1$. This
transformation can be mapped into the alternative equivalent form
$|x\rangle \to (-1)^{f(x)}|x\rangle$. It is this second form we shall
use below.  The aim in Grover's problem is to find $x_0$ by addressing
the oracle as few times as possible.  A classical search algorithm
would need the oracle to be addressed $O(M)$ times, whereas Grover's
algorithm requires only $O(\sqrt{M})$ calls of the oracle.

Consider an object composed of M pixels. As a first stage suppose that
each pixel is completely transparent, but that a single pixel $x_0$
induces a phase of $-1$, whereas all the other pixels induce a phase
of $+1$.  We are then exactly in the situation of Grover's algorithm,
and can determine which pixel induces the anomalous phase in
$\sqrt{M}$ passages through the object.

Now suppose that the object has a very small probability $\beta^2 \ll
1$ of absorbing a photon each time it passes through a pixel. Then the
probability that Grover's algorithm succeeds is $|\alpha|^{2
\sqrt{M}} \simeq \exp(-\beta^2 \sqrt{M})$.  On the other hand the
probability that an algorithm that addresses each pixel individually
succeeds without absorbing a single photon is $|\alpha|^{2 M} \simeq
\exp(-\beta^2 M)$.  Thus if $\beta^{-2} \ll M \ll \beta^{-4}$, one
can find the anomalous pixel using Grover's algorithm with vanishing
probability that a photon is absorbed, whereas addressing each
pixel individually would entail a large number of absorbed
photons. Note that if the number $M$ of pixels is larger than
$\beta^{-4}$, then Grover's algorithm will no longer work because the
photon will be absorbed before completion of the algorithm.

We now create an ``epsilon'' version of this problem by supposing that
the anomalous pixel $x_0$ induces a phase $e^{i\epsilon}$ relative to
all the other pixels. In this case we can either replace the call to
the oracle in Grover's algorithm by $n$ successive passages of the
photon through the pixels, with $n \epsilon = \pi$; or we can use
Fahri and Gutmann's continuous version of Grover's
algorithm\cite{FG}. In either case the total number of times the
photon must pass through the object is $N \simeq \sqrt{M} /
\epsilon$. The probability that the photon is absorbed is therefore
$P_{abs} \simeq \beta^2 \sqrt{M} / \epsilon$. Just as with the
limiting case, $\alpha_1=1$, $\alpha_2=1-2\epsilon$, of
absorption-free measurement, a violation of (\ref{E1}) seems at first
sight possible. However, this is only attainable by having $P_{abs}$
small, which means that $\beta^2 \ll \epsilon$, so again the
conditions for our inequality are not satisfied.

Qualitatively, the advantage of taking the object to be extremely
transparent is that quantum coherence can be maintained over many
passages of a photon through the object. This is illustrated not only
by the examples above, but also by the interferometric example at the
end of section \ref{three}, where, in the limit of extreme
transparency, it is advantageous to use a protocol in which the photon
passes many times through the object. The latter situation gave
absorptions consistent with our inequality (\ref{E1}), whereas in the
examples considered in this section $\beta$ and $\epsilon$ are of
equal (small) magnitude, so the inequality does not apply. It would be
interesting to find bounds which hold when both $\beta$ and $\epsilon$
tend to zero.

\section{Conclusion}\label{eight}

The possibility of absorption-free measurements \cite{EV}
suggests that one might be able to use the quantum character of light
to significantly improve the imaging of photosensitive objects, where
one wants to minimize the number of absorbed photons. However, as we
have just seen in the preceding section, discrimination of close
values of transparency by repeated trials of absorption-free
measurement leads to a bound on the number of absorbed photons
compatible with (\ref{E1}).

The conditions under which we have derived (\ref{E1}) are more general
than this, since absorption-free measurement uses one photon
at a time, whereas we allow protocols with arbitrary numbers of
photons.  Our result imposes an inescapable bound of order
$1/|\epsilon|^2$ on the mean number of absorbed photons. In most cases
this bound can be attained by simply sending a certain number of
photons through the object and counting the number of transmitted
photons. When $\alpha_1$ and $\alpha_2$ differ only in phase, an
interferometric set-up in which each photon can either pass through
through the object or take an alternative route also gives rise to a
mean number of absorbed photons of order $1/|\epsilon|^2$. However
when the probability that the object absorbs a photon tends to zero,
it is advantageous to use a modified protocol in which the photon
passes through the object many times.

Let us note parenthetically that using classical light as a source is
in general disadvantageous, particularly in the limit where the
object is very transparent. This seems to be because the photon number
is ill-defined for classical light, and this extra noise implies that
more photons must be sent through the object compared to a source with
well-defined photon number.

Of course, the problem of discriminating two close transparencies is a
specialized one. A more realistic problem would be that of putting
narrow limits on a transparency that can take a continuous range of
values. However, any quantum protocol that achieves the latter goal
must select from a number of measurement outcomes, and must therefore
be able to discriminate between pairs of outcomes for close
transparencies. Our $1/|\epsilon|^2$ bound therefore applies to such
protocols, with $\epsilon$ now being the standard deviation of the
estimated transparency. In general, counting the number of absorbed
photons will require of order $1/|\epsilon|^2$ photons to limit the
estimated transparency within a standard deviation of $\epsilon$.
Only in the case where the objects have the same probability of
absorbing a photon, but differ in the phase they induce, is an
interferometric protocol necessary.

Despite these rather negative conclusions for the single pixel case,
interferometric protocols can offer a significant advantage when the
task is to discriminate patterns of many pixels, and where all the
pixels are addressed collectively rather than individually. The
example of discrimination between a set of orthogonal patterns (added
to a constant background) shows that a $\log M$ gain is possible for
$M$ pixels, and Kent and Wallace have shown a gain of the same order
in the detection of defective pixels \cite{KW}. We have also
considered some special cases of absorption-free measurement and
searches with Grover's algorithm which do not meet the background
assumptions of our inequality (\ref{E1}) and are thereby able to
exceed the bounds it imposes. Finding further algorithms that can
answer questions about pixel arrays with low absorption cost seems an
interesting area for further exploration.

\section{Aknowledgements}
We thank Adrian Kent for helpful conversations and for drawing our
attention to the possibility of a logarithmic gain of quantum
multi-pixel algorithms compared to classical ones. We acknowledge
financial support by the European Science Foundation.  S.M. is a
research associate of the Belgian National Research Fund, he
acknowledges funding by the European Union under project EQUIP
(IST-FET program).


\end{document}